# Wave propagation dynamics inside a complex scattering medium by the temporal control of backscattered waves

Ye-Ryoung Lee[1,2,3], Wonjun Choi[1,2], Seungwon Jeong[1,2], Sungsam Kang[1,2], Dong-Young Kim[1,2], and Wonshik Choi[1,2,*]

[1]*Center for Molecular Spectroscopy and Dynamics, Institute for Basic Science, Seoul 02841, Korea*
[2]*Department of Physics, Korea University, Seoul 02855, Korea*
[3]*Institute of Basic Science, Korea University, Seoul 02841, Korea*
[*]*wonshik@korea.ac.kr*

**Abstract:** Shaping the wavefront of an incident wave to a complex scattering medium has demonstrated interesting possibilities, such as sub-diffraction wave focusing and enhancing light energy delivery. However, wavefront shaping has mainly been based on the control of transmitted waves that are inaccessible in most realistic applications. Here, we investigate the effect of maximizing the backscattered waves at a specific flight time on wave propagation dynamics and energy transport. We find both experimentally and numerically that the maximization at a short flight time focuses waves on the particles constituting the scattering medium, leading to the attenuation of the wave transport. On the contrary, maximization at a long flight time induces constructive wave interference inside the medium and thus enhances wave transport. We provide a theoretical model explaining this interesting transition behavior based on wave correlation. Our study provides a fundamental understanding of the effect of wave control on internal wave dynamics.

## Main text

Light waves propagating through complex media are scattered multiple times due to the refractive-index inhomogeneities, causing a loss of image information and the attenuation of wave energy transport. Studies have shown that the interference of the multiply-scattered waves can be manipulated by controlling the incident wavefront[1,2]. Using the wavefront shaping, spatial and temporal focusing were realized through a scattering layer[3–8], and transmission enhancement was demonstrated by coupling the incident wave to the eigenchannels of a transmission matrix[9–12]. Furthermore, a few noteworthy studies have explored the internal energy distribution of eigenchannels with analytic and numerical approaches[13–17], and experiments were conducted to reveal internal energy distribution using 2D waveguide[11,18,19] or fluorescent nanospheres[20]. While these studies have shown the great potential of wavefront shaping, they all relied on the monitoring and control of the transmitted waves in a steady state.

For most practical applications including deep-tissue bio-imaging, looking around the corner, and seeing through the foggy environment, it is possible to measure only the backscattered waves from the scattering medium. The transmitted waves are not allowed to measure because a detector cannot be placed either on the opposite side or within the medium. Therefore, using the backscattered waves as the beacons of wavefront shaping is crucial. An innate difference exists between the backscattered waves and the transmitted waves. Backscattered waves experience roundtrip while the transmitted waves make only a one-way trip through the scattering medium. This suggests that the effect of controlling the backscattered waves is more challenging to understand than that of the transmitted waves. Although adding the temporal degree of freedom by the time-gated measurement of the backscattered waves[21–23] and their temporal control[24] have reduced the complexity of exploiting the backscattered waves, the effect of wavefront shaping on general inhomogeneous scattering medium has remained to be explored.

This study aimed to address a key question regarding the effect of controlling backscattered waves on the internal field dynamics and wave energy transport. The wavefront shaping for maximizing the energy of the transmitted waves enhances wave transport through the scattering medium. Furthermore, minimizing the energy of the backscattered waves in the steady state enhances wave transport in the waveguide geometry or its equivalence[10]. This minimization strategy is ineffective in a thick scattering medium where waves can leak out of the control region. In this general situation, we explored if the maximization of backscattered waves at a specific flight time can either enhance or reduce the wave energy transport through the scattering medium.

Here, we investigate the wave propagation dynamics of the time-dependent reflection eigenchannels within a scattering medium. We conduct a finite-difference time-domain (FDTD) simulation to identify the eigenchannels with the largest eigenvalue (i.e., largest reflectance) for a specific target flight time. By injecting a light pulse into each time-dependent eigenchannel, we observe the temporal evolution of waves inside the scattering medium and find an interesting transition from wave focusing to diffusion with the increase of target flight time. At target flight times shorter than the transport mean free path, the eigenchannel tends to focus on particles constituting the scattering medium. This gives rise to strong backscattering, thereby attenuating wave transmission through the scattering medium. At longer target flight times, the eigenchannel is inclined to induce constructive wave interference inside the scattering medium, which leads to enhanced transmittance. We provide a theoretical model explaining the transition from attenuation to the enhancement of wave energy transport based on wave correlation. We experimentally validated this transition behavior with the increase in the target flight time, supporting the numerical simulation and theoretical model.

**Numerical simulation for observing internal field dynamics of eigenchannels**

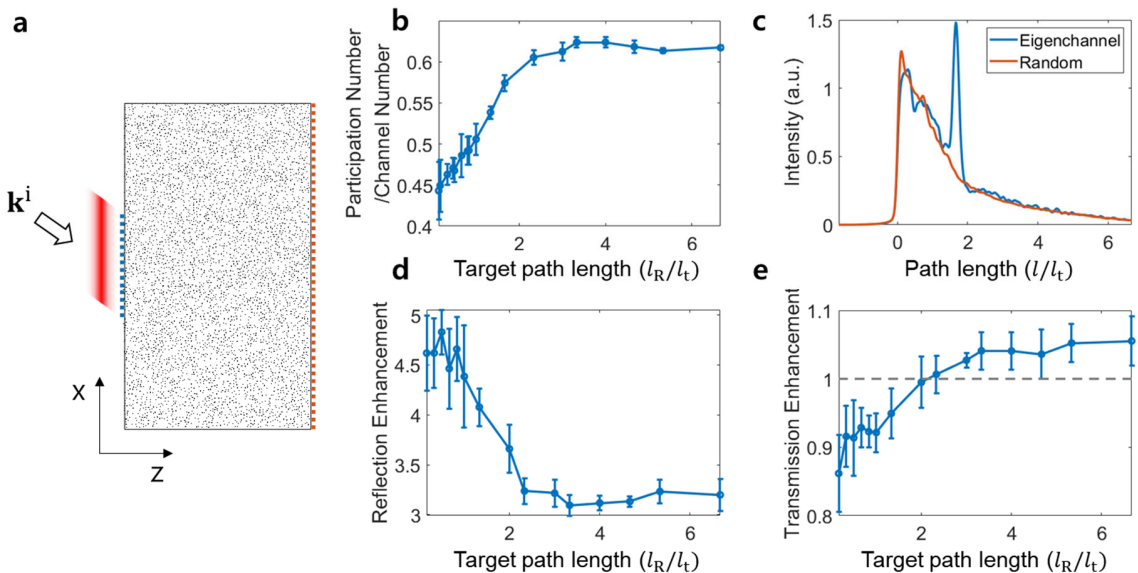

**Figure 1. Numerical simulation of time-gated reflection eigenchannels inside a complex scattering medium. a,** A two-dimensional scattering medium prepared for the simulation. The dashed blue line indicates the illumination and detection window for reflection measurements. The dashed red line indicates the detection window for transmission measurements. **b,** Normalized participation number depending on the target path length $l_R$ in the unit of $l_t$. **c,** Reflection intensities depending on the optical path length ($l$) when the incident wave is coupled to the time-gated reflection eigenchannel of $l_R = 50$ μm (blue) and when a random input is coupled (red). **d,** Maximum reflection enhancement achieved by the time-gated reflection eigenchannels depending on $l_R$. **e,** Transmission enhancement of the time-gated reflection eigenchannels when the incident wave is coupled to the eigenchannel with the largest eigenvalue for each target path length, $l_R$.

We conducted numerical simulations using the finite-difference time-domain (FDTD) method to observe the dynamics of the time-gated reflection eigenchannels inside a scattering medium. We numerically prepared a two-dimensional scattering medium comprised of absorption-free circular particles with a diameter and refractive index of 320 nm and 1.26, respectively (Fig. 1a). The particles were randomly distributed in a vacuum with a filling factor of 4.66%. The scattering and transport mean free paths at the center wavelength ($\lambda = 800$ nm) of the light source were calculated to be $l_s = 15$ μm and $l_t = 30$ μm, respectively. The width and thickness of the medium were 124 μm and 70 μm, respectively. Simulations were conducted for five different samples with similar scattering properties to ensure the reliability of the observations. A pulsed planar wave was injected through a 40 μm-width window in the middle of the scattering medium (dashed blue line in Fig. 1a), and its electric fields were computed within the scattering medium as a function of time. The pulse width of the light source was 16.7 fs, corresponding to a coherence length of $l_c = 5$ μm.

With this FDTD simulation platform, we computed the time-gated reflection matrices. For a planar wave with a specific incident transverse wave vector $\mathbf{k}_{in}$, the electric field $E_R(x, \tau_R; \mathbf{k}_{in})$ of the reflected wave was recorded at a position $x$ along the blue dashed line in Fig. 1a as a function of the flight time $\tau_R$. Using this information, we constructed a time-gated reflection matrix $\mathbf{R}(x, \tau_R; \mathbf{k}_{in})$, where $E_R(x, \tau_R; \mathbf{k}_{in})$ was assigned to the row and column indices of $\mathbf{R}$ corresponding to $x$ and $\mathbf{k}_{in}$, respectively. The time-gated reflection matrix at $\tau_R$ was expressed in terms of the optical path length $l_R = c\tau_R$, i.e. $\mathbf{R}(x, l_R; \mathbf{k}_{in})$, where $c$ is the average speed of light within the scattering medium. The reflection matrix was recorded for various $l_R$ with an interval of $\Delta l_R = 40$ nm. We performed the singular value decomposition of $\mathbf{R}(x, l_R; \mathbf{k}_{in})$ for each reflection matrix and identified its input eigenchannels $\mathbf{v}_j$ associated with the singular value $\sigma_j$. The singular values $\sigma_j$ of $\mathbf{R}$ were sorted in descending order with respect to the eigenchannel index $j$, and their associated eigenchannels $\mathbf{v}_j$ were indexed accordingly.

We examined the time-gated reflection matrices by calculating the participation number of each matrix, $M(l_R) = \langle \left( \sum_{j=1}^{N} \sigma_j^2 \right)^2 / \left( \sum_{j=1}^{N} \sigma_j^4 \right) \rangle$, where $N$ denotes the number of input channels. The participation number indicates the available degrees of freedom, and a stronger correlation effect is expected when the participation number is smaller[12]. Figure 1b shows the participation number of $\mathbf{R}(x, l_R; \mathbf{k}_{in})$ normalized by the number of output channels. The circles indicate the ensemble-averaged values of five different samples, and the error bars show the standard deviation. The participation number increases until the target path length ($l_R$) increases to approximately $2l_t$, suggesting that the wave correlation decreases as $l_R$ increases. Figure 1c shows reflection intensity depending on the optical path length $l$ of the reflected wave when the incident light is coupled to $\mathbf{v}_1$, the first eigenchannel of the time-gated reflection matrix with the largest singular value, measured at $l_R = 50$ μm (blue curve). A clear peak is observed at $l = l_R$ compared to the random input (red), confirming that the coupling to $\mathbf{v}_1$ raises the total intensity of the reflected wave at the target path length (see Supplementary Information Section I for other $l_R$s).

The maximum reflection enhancement, $\eta_R = \tau_{R,1}/\langle \tau_R \rangle$, achieved by the time-gated eigenchannel is plotted depending on the target path length in Fig. 1d. The reflection enhancement decreases until the target path length increased approximately up to $2l_t$, which agrees well with the observed decrement in the wave correlation with $l_R$ in Fig. 1b. Figure 1e shows the transmittance of the waves injected to $\mathbf{v}_1$ relative to that of the random input, which we define as transmission enhancement ($\eta_{tr}$). We evaluated the total transmittance through a 124 μm-width window (dashed red line in Fig. 1a) for the time-gated reflection eigenchannels of various target path lengths. For short target path lengths, the transmission is suppressed when coupling to $\mathbf{v}_1$. On the other hand, the transmittance is enhanced for the target path length longer than approximately $2l_t$. This observation that the transmittance experiences a transition from suppression to enhancement as the target path length increases contrasts with the fact that the reflectance is consistently enhanced for any target path length.

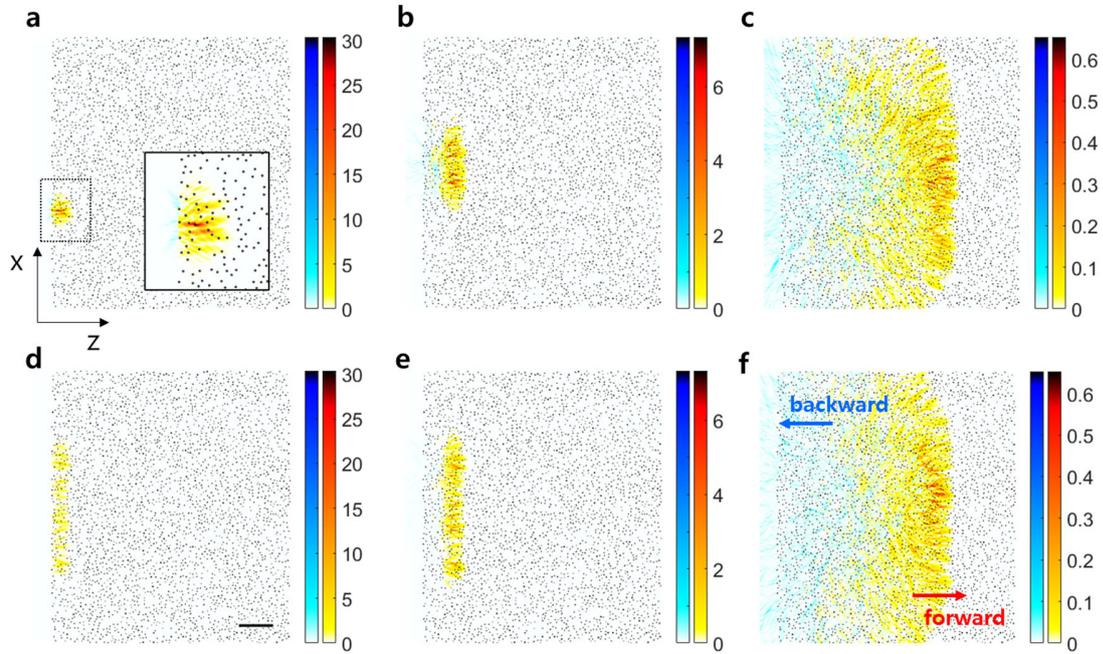

**Figure 2. Dynamics of the internal energy distribution of the time-gated reflection eigenchannel. a–c,** Internal energy distribution of the time-gated reflection eigenchannels for the target path lengths $l_R$ of 5 μm, 20 μm, and 100 μm, respectively, captured at the propagation time corresponding to half the target path length. The inset shows the magnified view of the region marked with a dotted rectangle. **d–e,** Internal energy distribution of the random input at the same propagation time as **a–c**, respectively. The intensity of the forward propagating waves is indicated in red, while that of the backward propagating waves is displayed in blue.

To elucidate the origin of the unique transition behavior of the transmittance, we investigated the dynamics of the internal energy distribution when the incident wave is coupled to the time-gated reflection eigenchannel. We recorded the complex field maps of the waves inside the scattering medium with time (see Supplementary Video I). We primarily investigated the internal field distribution of the eigenchannel with the largest eigenvalue for each reflection matrix and compared it with that of a random input. Forward and backward propagating waves were sorted out by taking positive and negative z-components of the wave vectors, respectively, obtained from the 2D Fourier transform of the complex field maps. Figure 2a–c shows the internal energy distribution of eigenchannels with the target path lengths $l_R$ of 5 μm, 20 μm, and 100 μm, respectively, captured at the propagation time of $\tau = l_R/2c$, corresponding to half the target path length. At this time, the waves have traveled half of their target path lengths. Figure 2d–f shows the distribution of the random input at the same propagation time as Fig. 2a–c, respectively, as a point of reference.

We observed the internal field distributions of the eigenchannels and noticed that the reflectance enhancing mechanism of the time-gated reflection eigenchannel differs based on the target path length. At a short target path length ($l_R = 5$ μm), the internal energy is focused on individual particles constituting the scattering medium as shown in the inset of Fig. 1a. In this case, the eigenchannel enhances reflectance by focusing waves on particles compared to the random input. The internal energy distribution of the eigenchannel at a long target path length ($l_R = 100$ μm) shows little difference from that of random input. As we shall show below, the eigenchannel enhances reflectance by increasing wave energy inside the scattering medium by inducing constructive interference of the multiply-scattered waves. The internal energy distribution of the eigenchannel for $l_R = 20$ μm shows transition characteristics between focusing and diffusion.

**Degree of wave focusing of the time-gated reflection eigenchannels**

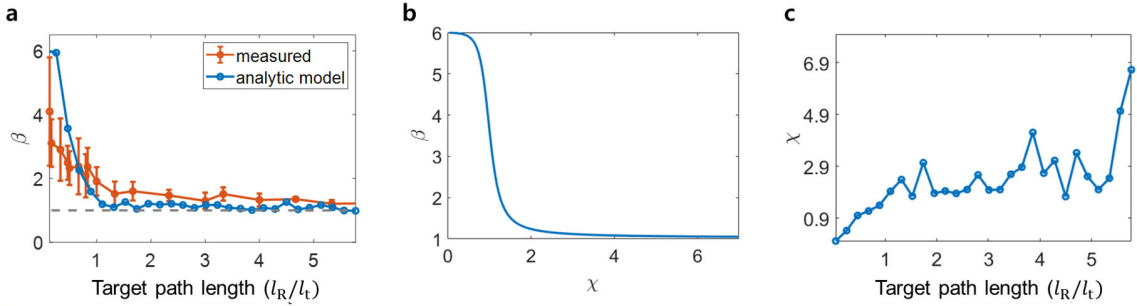

**Figure 3. Degree of wave focusing of the time-gated reflection eigenchannels. a,** Degree of focusing parameter $\beta$ depending on target path length $l_R$. Red dots: measured $\beta$ from the internal energy distribution at the propagation time corresponding to half the target path length. Blue dots: analytically derived $\beta$ from $\beta$ versus $\chi$ in **b** and $\chi$ versus $l_R$ in **c**. Gray line: reference line indicating $\beta = 1$. **b,** $\beta$ dependence on the governing parameter, $\chi = (\eta_B - 1)\langle\tau_B\rangle/(\eta_P - 1)\langle\tau_P\rangle$. **P** represents a sub-matrix describing multiple-scattered waves having interacted with particles at a specific depth range and **B** represents a sub-matrix for those multiple-scattered waves having no interaction with these particles. $\langle\tau_X\rangle$ denotes the average eigenvalue and $\eta_X \equiv \tau_{X,1}/\langle\tau_X\rangle$ corresponds to the enhancement factor of a given matrix **X**, where $\tau_{X,1}$ is the largest eigenvalue. **c,** $\chi$ depending on target path length $l_R$.

To quantitatively measure the transition behavior from focusing to diffusion depending on the target path length $l_R$, we define the degree of focusing parameter $\beta$ by the ratio of the energy enhancement at the particles ($\eta_P$) to the total energy enhancement ($\epsilon$) of the reflection eigenchannels at the depth range of interest:

$$\beta = \eta_P/\epsilon, \tag{1}$$

where $\beta$ indicates the effectiveness of the wave focusing on the particles when the incident wave is coupled to the reflection eigenchannel. When all the energy is focused on the particles present at the depth range of interest, $l_R/2 - \Delta l/2 \leq l \leq l_R/2 + \Delta l/2$, the energy at the particles is increased by a factor of $\eta_p = \epsilon/s$, where $s$ corresponds to total cross section ($1/l_s$) times depth range of interest ($\Delta l = l_c/2$), i.e. $s = \Delta l/l_s$. In this case, the focusing parameter becomes $\beta = 1/s$. In the opposite case when the focusing on particles does not occur at all, the focusing parameter is set $\beta = 1$.

We measured the wave focusing parameter $\beta$ directly from the internal energy distribution at the propagation time corresponding to half the target path length. We located the depth plane where the particle having the maximum intensity exists. After that, we obtained $\beta$ by measuring the total intensity enhancement ($\epsilon$) and enhancement of the region where particles exist ($\eta_p$). The measured $\beta$ (red circles in Fig. 3a) clearly shows the transition from focusing ($\beta > 1$) for the target path length shorter than $l_t$ to diffusion ($\beta \sim 1$) for a longer targe path length than $2l_t$.

The wave focusing is prominent at short target path lengths because the reflection signals from individual particles at a depth of interest contribute to the major fraction of the reflection matrix. For long target path lengths, both the ballistic and multiple-scattered waves from the individual particles in the target depth range are getting weaker than the multiple-scattered waves from other depths. To support this reasoning, we explain the transition from focusing to diffusion using an analytic model. Our previous study presented a framework for understanding the competing subprocesses and derived an analytic expression describing the way the eigenchannel coupling of the total process distributes input energy to individual subprocesses using the largest eigenvalue and the average eigenvalue of each subprocess[25].

This framework can be applied to explain the competition between multiple-scatted waves from particles at a specific depth range (***P***) and from the other particles at all the shallower depths (***B***) when the incident wave is coupled to a time-gated reflection matrix (***R***) of the total process, i.e., ***R*** = ***P*** + ***B***. The preferential coupling to ***P*** increases the multiply scattered waves having interacted with particles at a specific depth range, which leads to wave focusing on the particles. On the contrary, the preferential coupling to ***B*** increases multiple-scattered waves from all the shallower depths, thereby reducing the degree of wave focusing.

Two representative parameters considered in the previous study are the average eigenvalue $\langle \tau_X \rangle$ and the enhancement factor $\eta_X \equiv \tau_{X,1}/\langle \tau_X \rangle$ of a given matrix ***X***, where $\tau_{X,1}$ is its largest eigenvalue. The $\eta_X$ indicates the effectiveness of the wavefront control in maximizing the output of the corresponding process. We derived that $\chi \equiv \frac{(\eta_B - 1) \langle \tau_B \rangle}{(\eta_P - 1) \langle \tau_P \rangle}$ is a governing parameter determining the degree of coupling to each subprocess when the incident wave is coupled to the eigenchannel of the total process. If $\chi < 1$, reflection eigenchannels preferably couple light to the process ***P***, which works in favor of raising the wave focusing parameter $\beta$. Exploiting this framework, we can find the degree of coupling to ***P*** and ***B*** depending on $\chi$. Using this obtained degree of coupling and the estimated one-way energy enhancement of ***P*** and ***B*** from the FDTD simulation, we derived the relation between $\beta$ and $\chi$ (Fig. 3b, see Methods for detail). $\beta$ shows drastic change around $\chi = 1$ as expected.

To obtain the relationship between $\chi$ and $l_R$, we estimated $\langle \tau_B \rangle/\langle \tau_P \rangle$ for different target path lengths. We conducted FDTD calculations of the time-gated reflection intensities with and without the particles in the target depth range. $\langle \tau_P \rangle$ was determined by their difference in the flight time range corresponding to the target path length, and $\langle \tau_B \rangle$ was determined by reflection intensity at the corresponding flight time in the absence of particles in the target depth range (see Supplementary Information Section II). $\eta_P$ was estimated as the inverse of scattering probability ($\eta_P = 1/s$), which is the energy enhancement at the particles when all the incoming energy of the depth range is directed to interact with the particles. And $\eta_B$ was estimated as the measured reflection enhancement of eigenchannels with the longest path length, where the diffusion mechanism dominantly enhances the energy delivery. Combining all these estimations, we determined $\chi = \frac{(\eta_B - 1)\langle \tau_B \rangle}{(\eta_P - 1)\langle \tau_P \rangle}$ depending on target path length (Fig. 3c).

Eventually, we found $\beta$ depending on $l_R$ (blue circles in Fig. 3a). Derived $\beta$ clearly shows the transition from focusing ($\beta > 1$) for path length shorter than $l_t$ to diffusion ($\beta \sim 1$) for path length longer than $2\,l_t$. Furthermore, $\beta$ obtained from the analytic model (blue) agrees well with $\beta$ measured from the internal energy distribution (red), which demonstrates that the analytic model describes the physical origin behind the transition behavior of the time-gated reflection eigenchannel.

**Simple heuristic model explaining the transmittance of the reflection eigenchannels**

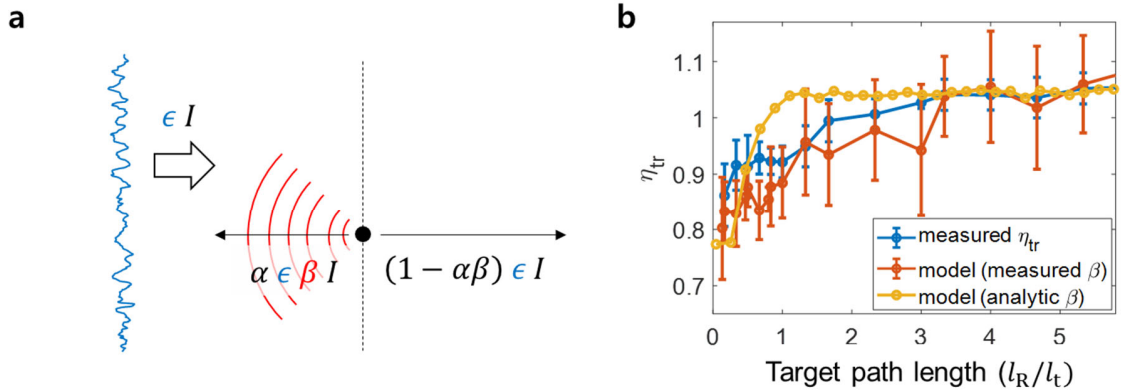

**Figure 4. Simple model explaining the transition behavior in transmission enhancement. a,** Model diagram describing the transition behavior in transmission enhancement. Black circular dot: a representative particle at a depth of interest indicated by the dashed line. $\alpha$: backscattering probability by the particles at a depth range of interest. Reflection intensity at a depth of interest is given by $\alpha \beta \epsilon I$, where $I$ denotes the average intensity at the depth for the random input. The resulting transmission is given by $(1 - \alpha \beta)\epsilon I$. **b,** Transmission enhancements predicted by the model using the measured $\beta$ (red) and analytically derived $\beta$ (yellow). Transmission enhancements measured directly from FDTD simulations are shown in blue.

We modeled the transition behavior in transmission introduced in Fig. 1e using the degree of focusing parameter $\beta$. The wave focusing on particles constituting the scattering medium gives rise to the backscattering by the particles, which in turn suppresses the transmittance. For a simplified picture, multiple target particles at a depth range of interest are represented as a single particle at a single depth plane in Fig. 4a. We define $\alpha$ as backscattering probability at a depth range of interest, which is given by the integral of differential cross-section, and $\epsilon$ as total energy delivery enhancement at the depth range of interest by eigenchannels. $\alpha$ was estimated to be 0.051 using the anisotropy factor $g = 0.5$ of particles used in the simulation, scattering mean free path $l_s = 15$ μm, and depth range of interest ($\Delta l$).

The eigenchannel coupling enhances reflection from $\alpha I$ of the random input to $\alpha \epsilon \beta I$ of the eigenchannel coupling (an arrow point to the left in Fig. 4a), inducing variation in transmittance from $(1 - \alpha)I$ of the random input to $(1 - \alpha \beta) \epsilon I$ of the eigenchannel coupling (an arrow pointing to the right in Fig. 4a). Therefore, the transmission enhancement by the time-gated reflection eigenchannels is given as

$$\eta_{tr} = \frac{(1-\alpha\beta)\epsilon}{1-\alpha}. \tag{2}$$

We predicted the transmission enhancement based on Eq. (2) using the $\beta$ and $\epsilon$ obtained from the internal energy distribution and the analytic model. The $\eta_{tr}$ obtained from the measured $\beta$ (red) and the analytic model (yellow) present the transition in transmission from suppression ($\eta_{tr} < 1$) to enhancement ($\eta_{tr} > 1$). We compared the $\eta_{tr}$ obtained from this model with the $\eta_{tr}$ measured directly from FDTD simulations in Fig. 1e. The simple model predicts the $\eta_{tr}$ reasonably well, supporting that the degree of wave focusing by the reflection eigenchannels is responsible for the transition of the transmittance enhancement.

The model assumed that the depth range of interest dominantly determined the transmission enhancement. For short target path lengths, the assumption is true since little diffusion occurs, and most energies are concentrated at the depth range of interest. For long target path lengths, substantial spatial spreads of waves along the depth exist. Still, the assumption can be true if $\epsilon$ is independent of depth and $\beta$ is approximately equal to 1 such that the depth range of interest represents other depths well, and this condition is indeed met (see Supplementary Information Section III). For target path lengths in the transition regime, the fraction of waves returning at the other depths may not follow the proposed model. However, the reasonable agreement between the direct observation and model suggests that the interaction at the target depth is mainly responsible for wave transport.

**Experimental measurement of the transmittance of the reflection eigenchannels**

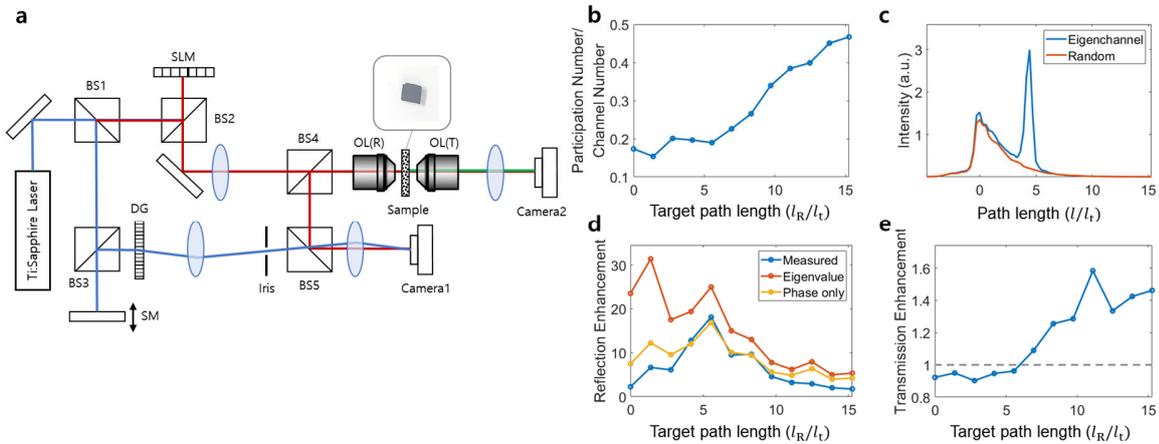

**Figure 5. Experimental observation of reflection eigenchannel-dependent transmittance. a,** Experimental

schematic. OL: objective lens; BS: beam splitters; SLM: spatial light modulator; DG: diffraction grating (an aperture was used to select the first-order diffracted wave); SM: path length scanning mirror. For clarity, red, blue, and green are used to indicate the sample path, the reference path, and the transmission path, respectively. **b,** Normalized participation number of the experimentally measured reflection matrix depending on the target path length $l_R$. **c,** Reflection intensities depending on the optical path length ($l$) when the incident wave was coupled to the time-gated reflection eigenchannel of $l_R = 4.4\, l_t$ (blue) and when a random input is coupled (red). **d,** Maximum reflection enhancement measured when the time-gated reflection eigenchannel was coupled for various target path lengths (blue). Red: Maximum reflection enhancement calculated from the corresponding time-gated reflection matrix. Yellow: Maximum reflection enhancement estimated for the phase-only approximation of eigenchannels. **e,** Transmission enhancement measured when the time-gated reflection eigenchannel with the largest eigenvalue is coupled depending on the target path lengths.

To validate our observation in the numerical simulation, we constructed an experimental system measuring the time-gated reflection matrix for various target path lengths $l_R$ (Fig. 5a). We experimentally shaped the wavefront of the incident wave to couple light to the reflection eigenchannel with the largest eigenvalue and measured its transmittance. A Ti:Sapphire laser (central wavelength of 780 nm and pulse width of 60 fs) was used as a light source for the femtosecond-scale time-resolved measurement. A phase-only spatial light modulator (SLM, X10468-02, Hamamatsu) was placed on the conjugate plane to the front surface of the sample. The incident angle of the input beam was scanned to obtain the reflection matrix by writing phase ramps on the SLM. The illumination area on the sample was $40 \times 40\ \mu m^2$. The total number of effective orthogonal channels for the numerical aperture of 0.4 of the objective lens was 1300. Complex field maps of the reflected waves were captured by camera 1 using off-axis holography, and transmission intensity images were taken by camera 2 over the field of view of $120 \times 120\ \mu m^2$. Time gating could be achieved by controlling the path length of the reference arm with a scanning mirror because we recorded images based on an interferogram. The inset in Fig. 5a shows a typical scattering sample used in the experiment. The sample was composed of randomly dispersed 100-nm silver powder in polydimethylsiloxane (PDMS) with a slab thickness of 185.7 μm. From the thickness, the average transmittance (10.6%), and the anisotropy factor of the silver powder ($g \sim 0$), both the scattering mean free path ($l_s$) and transport mean free path ($l_t$) were estimated to be 20.6 μm at a wavelength of 780 nm.

We constructed a time-gated reflection matrix $R(\mathbf{r}, l_R; \mathbf{k}_{in})$ using input phase ramps written on the SLM and a set of complex field maps of the reflected waves for each target path length $l_R$, where $\mathbf{k}_{in}$ represents the wavevector of the incident pulsed planar wave, and $\mathbf{r}$ corresponds to the spatial coordinates of the detection plane. We calculated the participation number of the reflection matrix for each target path length (Fig. 5b). Similar to the numerical simulation results, it increases steadily with the increase of the target path length. After that, we shaped the incident wave as a reflection eigenchannel with the largest eigenvalue of each reflection matrix and recorded the reflection and transmission images for each target path length. We only wrote the phase map of the eigenchannel on the SLM because SLM used in the experiment operated in phase-only mode. Figure 5c shows the reflection intensity depending on the path length $l$ when the incident light was coupled to the eigenchannel of the time-gated reflection matrix measured at $l_R = 4.4\, l_t$. A clear peak was observed at $l = l_R$ in the case of eigenchannel coupling compared to the random input (see Supplementary Information Section I for the case of other $l_R$s), which shows that the incident wave was successfully injected to the eigenchannel with the largest eigenvalue.

Fig. 5d (blue dots) shows the plot of the maximum reflection enhancement depending on the target path length. The reflection enhancement was increased until the target path length increased up to approximately

$5l_t$. The phase-only eigenchannel is not fully effective for short path lengths because the amplitude of the reflection eigenchannel is rather localized in this regime. In fact, the maximum reflection enhancement estimated by the measured matrix when the light was coupled to the phase-only approximation of the eigenchannels shows a similar tendency (yellow dots in Fig. 5d), supporting that this is an innate limitation of the phase-only SLM. The maximum reflection enhancement, $\eta_R = \tau_{R,1}/\langle\tau_R\rangle$, calculated from the measured matrix (red dots in Fig. 5d) shows a higher reflection enhancement in accordance with the smaller participation number in Fig. 5b. This suggests that the enhancement in this regime would increase if an SLM capable of controlling both phase and amplitude was used. When $l_R > 5l_t$, the reflection enhancement decreases as the target path length increases, showing a similar tendency to the numerical simulation.

Figure 5e shows the transmission enhancement ($\eta_{tr}$) of the reflection eigenchannel with the largest eigenvalue depending on the target path length. For short target path lengths, the transmission was suppressed when coupling the reflection eigenchannel. On the other hand, the transmission was enhanced when coupling the reflection eigenchannel for a target path length longer than $6l_t$. The transition behavior in transmission was observed experimentally, which exhibits the same trend as the simulation and the analytic model. The transmission suppression for short target path length is less prominent than results from the model and the simulations because the fidelity of the eigenchannel coupling is low due to the use of a phase-only SLM. The value of the target path length at the transition in the experiment is different from that of the FDTD simulation. The main difference may arise because the FDTD simulation was conducted in the 2D sample due to calculation capacity while the experiment was performed in the 3D sample. The different scattering properties of the samples used in the simulation and the experiment also contributed to the discrepancy.

Our study provides an intuition for important potential applications requiring efficient wave energy transport. Examples include laser therapy for skin diseases, photodynamic therapy, photothermal therapy, and optogenetics. In these practical applications, the direct monitoring of the waves inside the medium is not allowed because one cannot place a detector at a depth of interest. Therefore, it is crucial to find a way to exploit the backscattered waves as beacons to control the waves inside the medium. In this respect, our study presenting the link between the time-dependent backscattered waves and the wave energy transport made on their roundtrip will provide a clue. It will be an interesting next step to investigate the effect of absorbing objects such as blood vessels and melanocytes in skin tissues on the wavefront shaping of temporal backscattered waves and to seek conditions favorable for the wave energy transport to these target objects.

## Methods

**Details of FDTD simulation.** The FDTD method provides a perfect platform to investigate the effect of wave interference inside a scattering medium. We adopted the original algorithms used in our previous studies[24,25], which send a pulsed planar wave of arbitrary incidence angle and compute the complex field map of its propagation inside a scattering medium as a function of flight time. We conducted a numerical study for a two-dimensional medium because FDTD simulation requires significant resources. A disordered medium composed of absorption-free circular particles with a diameter and reflective index of 320 nm and 1.29, respectively, was placed in the x-z plane. The particles were randomly distributed in a vacuum with a filling factor of 4.66%. The scattering and transport mean free paths were calculated to be $l_s = 15$ μm and $l_t = 30$ μm, respectively, at the center wavelength of light source $\lambda = 800\ nm$. The width and thickness of the medium were 124 μm and 70 μm, respectively.

We sent an ultrashort pulse of electromagnetic wave polarized perpendicular to the x-z plane. The pulse width of the light source was 16.7 fs, corresponding to a coherence length of 5 μm. A 40 μm-width window in the middle of the scattering medium (dashed blue line in Fig. 1a) was illuminated. The incident wavevector $\mathbf{k}_{in}$ was scanned to cover input free modes within the numerical aperture of 0.59. Note that the temporal pulse front was parallel to the x-axis regardless of $\mathbf{k}_{in}$. Hence, all the angular waves can be sent simultaneously in the input plane. The cell size used in FDTD calculation was 10 nm, guaranteeing sufficient computation accuracy. For each $\mathbf{k}_{in}$, we computed wave propagation and recorded the reflected wave at the blue dashed line in Fig. 1a as a function of flight time. A time-gated reflection matrix was constructed from the recorded maps and identified the time-gated reflection eigenchannels. Thereafter, we calculated the field maps of each eigenchannels inside the scattering medium, which is shown in the Fig. 2. The transmission field maps of each eigenchannels were calculated at the red dashed line in Fig. 1a.

**Derivation of the relation between $\beta$ and $\chi$.** We explain the derivation of the relation between $\beta$ and $\chi$ using the obtained degree of coupling and the estimated one-way energy enhancement of $\boldsymbol{P}$ and $\boldsymbol{B}$ from the FDTD simulation.

The degree of coupling can be defined as the squared correlation between the eigenchannel of the total process and that of the subprocess (probability), $\alpha_X^m = |\langle \mathbf{v}_{X,1} | \mathbf{v}_{R,1} \rangle|^2$. From the previous study[25], we derived the degree of coupling to each subprocess as

$$\alpha_P^m \approx \frac{1}{2} + \frac{1}{2} \frac{\frac{1}{\chi}-1}{\sqrt{\frac{4}{N}+\left(\frac{1}{\chi}-1\right)^2}}. \tag{3}$$

$$\text{and } \alpha_B^m \approx \frac{1}{2} + \frac{1}{2} \frac{\chi-1}{\sqrt{\frac{4}{N}+(\chi-1)^2}}, \tag{4}$$

where $\chi = \frac{(\eta_B - 1)\langle \tau_B \rangle}{(\eta_P - 1)\langle \tau_P \rangle}$.

Using the obtained degree of coupling to $\boldsymbol{P}$ and $\boldsymbol{B}$, we can extract $\epsilon$ and $\eta_p$ as

$$\epsilon \approx (\eta_B^{in} - 1)\alpha_B^m + 1 \tag{5}$$

$$\text{and } \eta_P \approx (\eta_P^{in} - 1)\alpha_P^m + 1, \tag{6}$$

where $\eta_P^{in}$ and $\eta_B^{in}$ denote enhancement factors on the way into the depth range (one-way) when coupling the eigenchannels of $\boldsymbol{P}$ and $\boldsymbol{B}$. For intuitive understanding, $\alpha_X^m$ represents the degree of resemblance of the eigenchannel of the total process with the eigenchannel of subprocess $\boldsymbol{X}$. Hence, the enhancement of the process $\boldsymbol{X}$ is approximately the summation of $\eta_X^{in}$ times the probability ($\alpha_X^m$) and 1 (no enhancement) times $1 - \alpha_X^m$ (the rest). The one-way enhancement factor of $\boldsymbol{P}$ ($\eta_P^{in}$) can be estimated the same as the roundtrip enhancement factor ($\eta_P$) because energy enhancement at particles directly enhances reflection. We approximated the one-way enhancement factor of $\boldsymbol{B}$ ($\eta_B^{in}$) as the measured transmission enhancement of the eigenchannel with the longest path length, which is dominantly determined by the diffusion mechanism. Using these parameters, we obtained the scattering probability enhancement from the definition, $\beta = \eta_P/\epsilon$.

**Acknowledgment**
This work was supported by the Institute for Basic Science (IBS-R023-D1), the National Research Foundation of Korea (NRF) grant funded by the Korea government (MSIT) (NRF-2021R1C1C2008158), and the POSCO Science Fellowship of POSCO TJ Park Foundation.